\documentclass{getwriting}

\usepackage[superscript, biblabel]{cite}
\usepackage[T1]{fontenc}
\usepackage{interval}
\usepackage{multirow}
\usepackage{comment}

\usepackage{mdframed}
\mdfdefinestyle{boxstyle}{%
linewidth=1.5pt,%
frametitlerule=false,%
innertopmargin=\topskip,
}
\mdtheorem[style=boxstyle]{mybox}{Box}

\title{\textbf{Original research article:} Is checking for sequential positivity violations getting
you down? Try sPoRT!}
\author[1]{Arthur Chatton \orcidlink{0000-0002-0018-5899} \thanks{Contact: arthur.chatton@umontreal.ca}}
\author[2,3]{Michael Schomaker}
\author[4,5]{Miquel-Angel Luque-Fernandez}
\author[6,7,8]{Robert W. Platt}
\author[1,8,9]{Mireille E. Schnitzer}
\affil[1]{\footnotesize Faculté de Pharmacie, Université de Montréal, Montréal, QC, Canada}
\affil[2]{\footnotesize Department of Statistics, Ludwig-Maximilians University Munich, Munich, Germany}
\affil[3]{\footnotesize Centre for Infectious Disease Epidemiology and Research, University of Cape Town, Cape Town, South Africa}
\affil[4]{\footnotesize Department of Statistics and Operations Research, University of Granada, Granada, Spain}
\affil[5]{\footnotesize Department of Epidemiology and Population Health, London School of Hygiene and Tropical Medicine, London, UK}
\affil[6]{\footnotesize Center for Clinical Epidemiology, Lady Davis Institute, Jewish General Hospital, Montr\'eal, QC, Canada}
\affil[7]{\footnotesize Department of Pediatrics, McGill University, Montr\'eal, QC, Canada}
\affil[8]{\footnotesize Department of Epidemiology, Biostatistics, Occupational Health, McGill University, Montr\'eal, QC, Canada}
\affil[9]{\footnotesize D\'epartement de M\'edecine Sociale et Pr\'eventive, Universit\'e de Montr\'eal, Montr\'eal, QC, Canada}

\begin{document}

\maketitle

\begin{abstract}
Background: Sequential positivity is often a necessary assumption for drawing causal inferences, such as through marginal structural modeling. Unfortunately, verification of this assumption can be challenging because it usually relies on multiple parametric propensity score models, unlikely all correctly specified. Therefore, we propose a new algorithm, called  \textit{sequential Positivity Regression Tree} (sPoRT), to overcome this issue and identify the subgroups found to be violating this assumption, allowing for insights about the nature of the violations and potential solutions.

Methods: We present different versions of sPoRT based on either stratifying or pooling over time under static or dynamic treatment strategies. \textcolor{black}{This methodological development was motivated by} a real-life application \textcolor{black}{of the impact of the timing of initiation of HIV treatment} with and without smoothing over time, \textcolor{black}{which we also use to demonstrate the method}.

Results: The illustration of sPoRT demonstrates its easy use and the interpretability of the results for applied epidemiologists. Furthermore, an \texttt{R} notebook showing how to use sPoRT in practice is available at github.com/ArthurChatton/sPoRT-notebook.

Conclusions: The sPoRT algorithm provides interpretable subgroups violating the sequential positivity violation, allowing patterns and trends in the confounders to be easily identified. We finally provided practical implications and recommendations when positivity violations are identified.\\


\end{abstract}

\section{Introduction}\label{intro}

Causal inference is a two-step process.\cite{Petersen_vanderLaan_2014} First, we define a causal parameter of interest in a given population based on counterfactual data. For instance, \textcolor{black}{parameters in }a marginal structural model (MSM) is a popular choice to represent longitudinal treatment effects. Based on a causal model representing knowledge and assumptions about the data-generating process, one can determine whether the counterfactual scientific question of interest (represented by the causal parameter) can be written as a statistical parameter of only the observed data. Next comes the estimation: one uses an estimator of the statistical parameter defined in the first step. Numerous estimators were developed for parameters defined to represent treatment effects in longitudinal settings, notably based on the g-formula,\cite{Robins_1986, Bang_Robins_2005} inverse probability weighting,\cite{Robins_2000} and doubly robust estimators, such as the longitudinal targeted maximum likelihood estimator.\cite{Petersen_Schwab_2014, Schnitzer_2014} Tutorials of various levels of technicality on these estimators are available.\cite{Daniel_2013, Schomaker_2019, Naimi_Cole_Kennedy_2016, Hernan_Robins_2020} 

While causal effects in longitudinal settings were defined and estimators proposed by Robins in the mid-80s,\cite{Robins_1986} later works investigated how to deal with common statistical challenges in practice. One major issue in longitudinal causal inference is violations (or near-violations) of the positivity assumption, which requires that all possible confounder patterns may be observed following all relevant treatments at all time points.\cite{Orellana_2010, Jensen_2024} \textcolor{black}{Such violations can be either structural or practical, and only expert domain knowledge can distinguish them \cite{Petersen_2012, Zhu_2021}. The former is due to an underlying phenomenon explaining why a particular subgroup of individuals is always (un)treated at or from a particular time-point. The latter occurs by chance only and is exacerbated in small and high-dimensional samples. With structural violations, identification bias will be present even without estimation bias \cite{Lundberg_2021}. In other words, bias due to targeting a causal parameter differing from the target one will be present even when there is no bias in the estimation of the statistical parameter. In contrast, practical violations only challenge the estimation process. In this last case, } model smoothing \textcolor{black}{(i.e., removing any kind of stratification)} allows for a relaxation of this requirement \cite{Cole_Hernan_2008}. Statistical covariate reduction methods can also be helpful.\cite{Schnitzer_2020} To diagnose the presence of sequential positivity (near-)violations, one may identify covariate-defined subgroups with null (or nearly null) estimated probabilities of some intervention trajectory's level. Thus, the diagnosis of positivity violations is complicated because positivity must be checked across all time points and may rely on many propensity score models \textcolor{black}{ fitted with increasingly complex covariate histories on decreasing sample sizes}, unlikely all correctly specified.\cite{Schomaker_2019, Rudolph_2022} This is a direct extension of the positivity assumption in single time-point treatment settings.\cite{Petersen_2012}  

The \textit{Positivity Regression Trees} (PoRT) algorithm\cite{Danelian_2023} is a regression tree-based explanatory tool to check positivity in single time-point treatment settings that require neither assumptions about the data-generating process nor specification of parametric models. In this study, we propose an extension of the PoRT method to longitudinal settings -- sequential PoRT or ``sPoRT" -- to identify the time-updated subgroups of individuals (defined by baseline and time-dependent covariate patterns), which lack data support for some category of treatment. We demonstrate the potential of this approach by reanalyzing a study investigating the effect of delaying HIV antiretroviral therapy (ART) among children in Southern Africa.\cite{Schomaker_2019}

\section{Motivating longitudinal data example}\label{illus}

ART is effective for reducing infant mortality,\cite{Violari_2008, Edmonds_2011} causing the WHO to recommend immediate ART initiation in all HIV-positive children as of 2015.\cite{WHO_2016} 

Schomaker \textit{et al.} investigated the effect of different ART initiation rules on the child's growth measured by height-for-age z-scores (HAZ) in West and Southern Africa using the observational International epidemiological Databases to Evaluate AIDS (IeDEA) in sub-Saharan Africa.\cite{Schomaker_2019, Egger_2012, Schomaker_2013, Schomaker_2016, Schomaker_2017} They highlighted the severity of positivity violations encountered in their analysis. We reanalyse a component of this study to demonstrate how PoRT can be used to identify sequential positivity violations. 

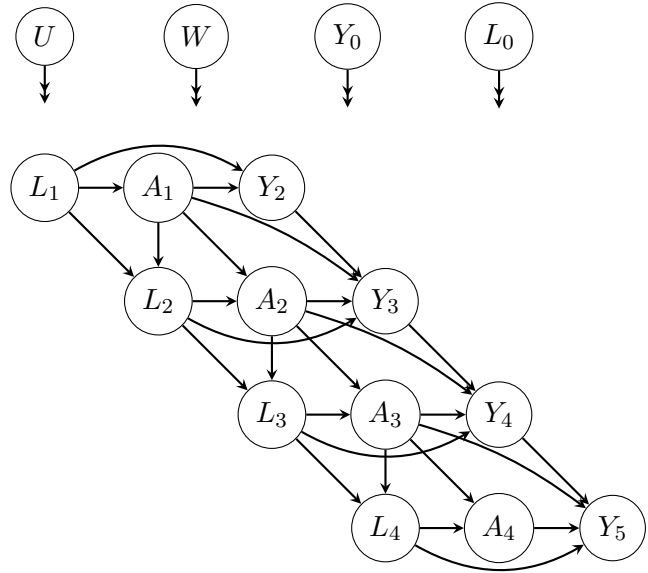
\begin{figure}[htb]
\centering
\begin{tikzpicture}
    \node[circle, draw](L1) at (0,4.5){$L_1$};
    \node[circle, draw](A1) at (1.5,4.5){$A_1$};
    \node[circle, draw](Y1) at (3,4.5){$Y_2$};
    \node[circle, draw](L2) at (1.5,3){$L_2$};
    \node[circle, draw](A2) at (3,3){$A_2$};
    \node[circle, draw](Y2) at (4.5,3){$Y_3$};
    \node[circle, draw](L0) at (6,6.5){$L_0$};
    \node[circle, draw](Y0) at (4,6.5){$Y_0$};
    \node[circle, draw](W) at (2,6.5){$W$};
    \node[circle, draw](U) at (0,6.5){$U$};
    \node[circle, draw](L3) at (3,1.5){$L_3$};
    \node[circle, draw](A3) at (4.5,1.5){$A_3$};
    \node[circle, draw](Y3) at (6,1.5){$Y_4$};
    \node[circle, draw](L4) at (4.5,0){$L_4$};
    \node[circle, draw](A4) at (6,0){$A_4$};
    \node[circle, draw](Y4) at (7.5,0){$Y_{5}$};
    \coordinate[shift={(0mm,-5mm)}] (n) at (W.south);
    \coordinate[shift={(0mm,-5mm)}] (m) at (U.south);
    \coordinate[shift={(0mm,-5mm)}] (o) at (L0.south);
    \coordinate[shift={(0mm,-5mm)}] (p) at (Y0.south);

    \draw[->, thick, >=stealth]
    (L1) edge (A1)
    (L1) edge (L2)
    (A1) edge (L2)
    (A1) edge (Y1)
    (A1) edge (A2)
    (A1) edge [bend left=10] (Y2)
    (L2) edge (L3)
    (A2) edge (L3)
    (Y2) edge (Y3)
    (A2) edge (A3)
    (A3) edge (Y3)
    (A2) edge [bend left=10] (Y3)
    (L3) edge (L4)
    (A3) edge (L4)
    (Y3) edge (Y4)
    (A3) edge (A4)
    (A3) edge [bend left=10] (Y4)
    (A4) edge (Y4)
    (L2) edge (A2)
    (L3) edge (A3)
    (L4) edge (A4)
    (Y1) edge (Y2)
    (A2) edge (Y2)
    (L1) edge[bend left=30] (Y1)
    (L2) edge[bend right=30] (Y2)
    (L3) edge[bend right=30] (Y3)
    (L4) edge[bend right=30] (Y4);
    \draw[->>, thick, >=stealth]
    (W) edge (n)
    (U) edge (m)
    (L0) edge (o)
    (Y0) edge (p);
    
\end{tikzpicture}
\caption{Simplified posited data-generating process for the first time points. Double-headed arrows indicated an effect on all downstream nodes. $A_t; t=1,2,3,4$ are the treatment status at each time point, $L_t; t=0,1,2,3,4$ are the time-varying confounder sets, $Y_t; t=0,2,3,4,5$ are the repeatedly measured outcomes. $W$ is the baseline confounder set, and $U$ are potential unmeasured confounders. \label{dag}}
\end{figure}

Our reanalysis includes data from South Africa, Malawi, and Zimbabwe. HIV-positive children aged 12-59 months at the first measured contact with a health care facility (which is the baseline) who acquired the virus before or at birth or during breastfeeding were included. To be eligible for the analysis, children had \textcolor{black}{to have at least one ART-naive visit}, complete data at enrolment (baseline), and at least one follow-up clinic visit. In total, 2352 children were included in the analysis. Data were recorded at 0, 1, 3, 6, 9, and 12 months after enrolment ($t=0,1,2,3,4,5$ respectively). At nine months, 1893 children were still followed. The most relevant parts of the posited causal structure from the initial study are depicted in Figure~\ref{dag}. More details regarding censoring and relevant baseline unmeasured confounders are given in the initial study.\cite{Schomaker_2019} The outcome $Y_t$ is HAZ measured at time $t$; the treatment $A_t$ refers to whether ART was taken at time $t$; the time-varying confounders set $L_t$ includes CD4 cell count, CD4\% (\textit{i.e.}, CD4/white blood cell counts $\times$ 100), and clinical stage, approximated by weight for age z-score (WAZ);\cite{Schomaker_2013} the time-fixed confounders set $W$ includes biological sex, age, and year of treatment initiation. $Y_0$ is the baseline HAZ, and it is included in the adjustment set.

\section{Treatment and treatment strategy}

\begin{figure}[htb]
    \centering
    \includegraphics[width=\textwidth]{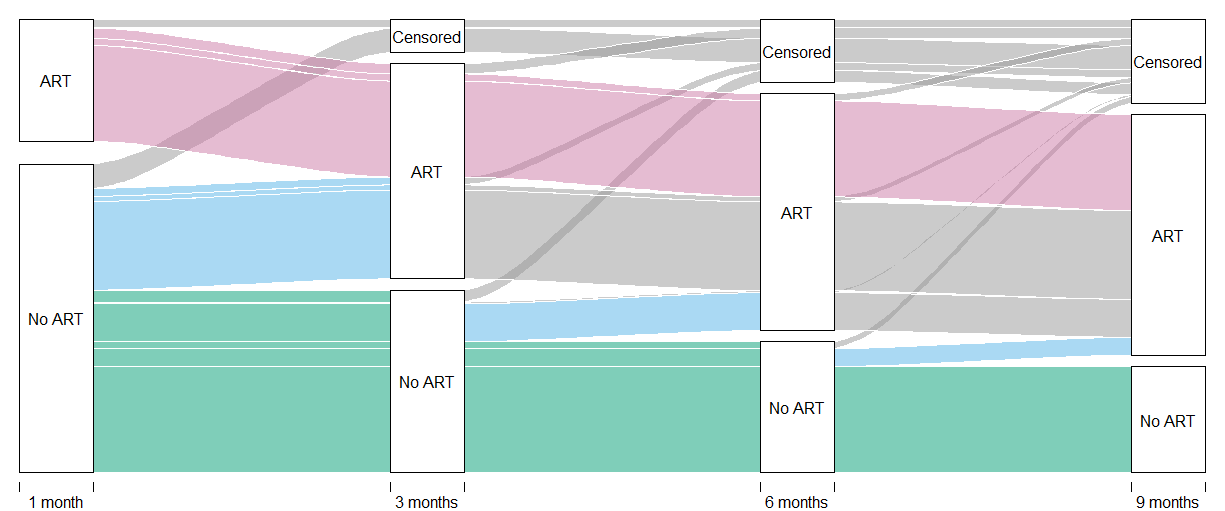}
    \caption{Switches between antiretroviral therapy (ART) usage. Immediate and no initiation are highlighted in magenta and green, respectively. Initiation is highlighted in black. \label{alluviate}}
\end{figure}

Time-varying treatment complicates the estimand definition. Furthermore, data support may vanish over time due to the potentially large number of possible treatment trajectories. Focusing on MSM parameters,\cite{Petersen_2014} if treatment can change at any of $T$ time-points, there can be as many as $2^T$ trajectories. In our example, the treatment, which represents ART initiation, is monotone such that $A_t=1$ when $A_{t-1}=1$ (see Figure \ref{alluviate}, magenta flow).

Treatment strategies can be used in longitudinal settings to define relevant causal contrasts flexibly. A treatment strategy is a rule that assigns the treatment at each time $t$ of follow-up.\cite{Smith_2022} To avoid confusion with observed treatment, we will now use the term \textit{intervention rule} for treatment strategy. A \textit{dynamic} rule depends on the 
value of one or more 
covariates. Otherwise, the rule is \textit{static}. Common examples of static rules are “always treat” and “never treat”. Dynamic rules are sometimes more realistic or even essential in practice. For instance, when some patients develop a contraindication to a specific treatment, they can no longer receive it and must switch to another one, which any realistic intervention rule must respect.\cite{Schnitzer_Platt_Durand_2020}  We denote an intervention rule at time $t$ by $d_t$.

We considered two static rules: immediate ART initiation and continued usage (rule 1, $d_t=1$ for $t=1,2,3,4$) or no ART initiation at any time (rule 2, $d_t=0$ for $t=1,2,3,4$). We also considered two dynamic intervention rules consisting of deferring ART initiation until a particular CD4 threshold is reached (similar to WHO's 2006 and 2010 guidelines\cite{Schomaker_2017}). The rules are to start treatment ($d_t=1$) when (i) CD4 count < 750 or CD4\% < 25\% (rule 3), or (ii) CD4 count < 350 or CD4\% < 15\% (rule 4). For these dynamic rules, we can write $d_t(L_t)$ since the rules are a function of the time-updated covariates. Of note, ART is always continued once initiated to respect the observed treatment monotonicity. In this demonstration, we do not diagnose sparsity in the censoring process, although this would be possible.

\section{The PoRT algorithm for single time-point treatment settings}\label{port}

First, consider \textcolor{black}{only the first} time-point \textcolor{black}{with the following data structure: $Y_2$} is the outcome, \textcolor{black}{$A_1$} is the binary treatment, and \textcolor{black}{$X=(L_0,L_1,W)$} is the confounder set. \textcolor{black}{For ease of notation, we remove the subscripts hereafter.} The PoRT algorithm runs a succession of regression trees, where the treatment is regressed on one or more confounders.\cite{Danelian_2023} An overview of the algorithm is as follows. First, fit individual trees by regressing the treatment on each confounder (\textit{i.e.} one tree per confounder). Second, search through each tree for nodes (\textit{i.e.}, subgroups of individuals) with an extreme probability of receiving a treatment level according to two user-defined hyperparameters ($\alpha$ and $\beta$, see Table \ref{tab:port}). Third, record the problematic subgroups identified in the second step and remove the related covariate(s) from further consideration. After the third step, repeat the above three steps, except this time, increase the number of covariates in each tree by one. The algorithm stops once the user-supplied value of the subgroups' allowed complexity (hyperparameter $\gamma$, Table \ref{tab:port}) is exceeded. 

The hyperparameter $\alpha \geq 0$ controls the size of identified subgroups by excluding the smallest ones. Positivity violations in very small subgroups may not impact the estimation when the model smooths over the subpopulation.\cite{Petersen_2012, Léger_2022} The hyperparameter $\beta$ defines the desired level of positivity, such that $P(A=a|\textcolor{black}{X})\leq\beta$ is considered a violation. It represents a threshold where the treatment probability becomes too extreme for the treatment effect in a subgroup to be supported by the data.\cite{DAmour_2021} Recently, Gruber \textit{et al.} proposed a new optimal bound of $5/(\sqrt{n} \ln{n})$ to truncate the propensity score,\cite{Gruber_2022} which can be applied as the default value of $\beta$. The last hyperparameter, $\gamma$, represents the maximum number of covariates used to build a regression tree. For instance, when $\gamma$ is set to two, the resulting subgroups are defined by a maximum of two covariates (\textit{e.g.}, male patients older than 70 years). Its optimal value depends on the treatment allocation mechanism's complexity. According to the positivity assumption, \textcolor{black}{$X$} should include all variables (confounders or not) that are part of the adjustment set used for future estimation.\cite{Westreich_2020} 
Furthermore, clinically meaningful violations may be missed when the full set \textcolor{black}{$X$} is used to build a tree.\cite{Danelian_2023}

PoRT's strength relies on its usage of regression trees. Regression trees are one of the most interpretable data-driven methods,\cite{Breiman_2001} making their output ideal for obtaining clinically interpretable subgroups defined by covariate values. They also can adapt to non-linear functional forms and higher-order interactions between covariates \textcolor{black}{without \textit{a priori} specification}.\cite{Bi_2019} Trees' weaknesses are their tendency to overfit and the greedy categorization of continuous variables. PoRT overcomes the first issue by fitting the simplest regression trees first and increasing their complexity sequentially through $\gamma$. When applying PoRT, the authors suggested manually categorizing the continuous confounders using meaningful thresholds since the interpretability of the subgroups is key,\cite{Danelian_2023} which also circumvents the second issue. \textcolor{black}{Categorization into multiple classes, each with data support, should be preferred over dichotomization to limit the loss of information.}

\begin{table*}[hb]
\scriptsize
\begin{center}
\caption{PoRT main hyperparameters\label{tab:port}}%
\begin{tabular}{clc}
\toprule
Hyperparameter & Definition & Suggested default values\textsuperscript{1}\\
\midrule
$\alpha$ & Minimal proportion of the whole sample size to consider a problematic subgroup & 0, 0.05, 0.1\\
$\beta$  & Threshold defining positivity violations: $P(A=a|W) > \beta,$ for all relevant $a$ & 0.01, 0.05, $5/(\sqrt{n} \ln{n})$\\
$\gamma$ & Maximal number of covariates in a tree & 2, 3, all \\
\hline
\multicolumn{3}{l}{Notation: $A$, Treatment; $n$, sample size; \textcolor{black}{$X$}, covariates \textcolor{black}{set}. $^{1}$ Must be based on domain expertise.}
\end{tabular}
\end{center}
\end{table*}

\section{sPoRT for interventions at multiple time points}

Sequential positivity implies that all individuals in the population have a non-zero probability of continuing to follow a given assigned intervention strategy at each time point, given that they have followed it thus far.\cite{Schomaker_2019} However, the way one may wish to check sequential positivity may vary depending on \textcolor{black}{whether one is estimating time-smoothed propensity scores for treatments pooled over time, or using time-stratified propensity scores of time-specific treatments}. Rudolph \textit{et al.} illustrated the advantages and pitfalls of different levels of smoothing with respect to positivity.\cite{Rudolph_2022} Briefly, stratifying (i.e., no smoothing) over time and/or treatment patterns improves the flexibility of the estimation when the model parameters vary significantly over time or across treatment histories. However, this stratification increases the number of models to fit, while reducing the amount of data available for their fit. Practical positivity violations (or data sparsity) are more likely to occur, yielding either estimation issues or the strong assumption of a correct extrapolation over missing strata.\cite{Léger_2022} \textcolor{black}{Smoothing} over time-points and/or treatment histories increases the stability of the estimation by positing a model that shares its parameters over time or across different treatment histories. However, smoothing over heterogeneous parameters may cause bias, even if the precision increases.\cite{Rudolph_2018} 

Our sPoRT algorithm can be used to check the sequential positivity assumption under either stratification or smoothing over time. If stratification is desired, the PoRT process described in Section \ref{port} is applied to each time point independently of each other, using a data subset depending on the intervention rule being investigated. The analyst must specify the time-varying confounders, the time-varying rules indicators, the hyperparameters, and any additional subsetting used for the estimation. 
The sPoRT algorithm can estimate the relevant treatment probabilities according to the observed treatment pattern (monotone or non-monotone) and the intervention rule type (Table \ref{tab:probamod}). Briefly, observed treatment monotonicity requires subsetting of individuals who have not initiated (or ceased) the treatment at the previous time. \textcolor{black}{For dynamic intervention rules, sPoRT assesses whether individuals who, under the specified rule, are theoretically expected to receive (or not receive) treatment at time $t$, are indeed treated (or untreated, respectively).} However, a common occurrence is that the data do not support such a stratification over treatment history due to decreasing sample sizes over time. In that case, one can smooth over treatment history, assuming the simplified model and covariates adjustment are appropriate \cite{Rudolph_2022}. \textcolor{black}{We posited in Figure \ref{dag} that the full confounders and ART histories until time $t$ do not impact ART initiation at time $t$. Otherwise, using propensity scores that smooth over these histories might invalidate the sequential conditional exchangeability assumption. When several practical violations that cannot be handled with a dynamic rule are found with a time-stratified sPoRT, one may once again face a tradeoff between non-positivity (not smoothing) and non-exchangeability (smoothing). Therefore, sPoRT may help gain an understanding, for a given rule, of how far into follow-up we can go without relying heavily on such smoothing assumptions.}

\begin{table}[htb]
\begin{center}
\caption{Illustration of typical probabilities modeled post-baseline in sPoRT using two time-points\label{tab:probamod}}%
\begin{tabular}{p{3cm}p{3cm}p{7cm}}
\toprule
Rule & Pattern of $A$ & Probability modeled\\
\midrule
Static & Monotone & $P(A_2=1 | A_1=0,L_2,W, d_2=1) > \beta$\\
&& $P(A_2=0 | A_1=0,L_2,W, d_2=0) > \beta$\\
&&\\
                        & Non-monotone & $P(A_2=1 | A_1,L_2,W, d_2=1) > \beta$\\
                        && $P(A_2=0 | A_1,L_2,W, d_2=0) > \beta$\\
                        &&\\
                        &&\\
Dynamic & Monotone & $P(A_2=1 | A_1=0, L_2, W, d_2(L_2)=1) > \beta$\\

                        && $P(A_2=0 | A_1=0, L_2, W, d_2(L_2)=0) > \beta$\\
                         &&\\
                        & Non-monotone & $P(A_2=1 | A_1, L_2, W, d_2(L_2)=1) > \beta$\\
                        && $P(A_2=0 | A_1, L_2, W, d_2(L_2)=0) > \beta$\\
\hline
\multicolumn{3}{p{15cm}}{Notation: $A_t$, treatment; $d_t(L_t)$, indicator of whether the rule recommends treatment; $L_t$, time-varying confounders; $W$, baseline confounders; $\beta$, positivity bound. Subscript indicates time-point. 
Note: Smoothing over treatment history is applied in the above. Different subsetting can be used depending on the specific context. Monotone refers to a non-decreasing sequence, without loss of generality.}
\end{tabular}
\end{center}
\end{table}

\section{Application results}

We ran sPoRT with stratification on time with the following hyperparameters: $\alpha=0.05$, $\beta=5/(\sqrt{n_t} \ln n_t)$, where $n_t$ is the sample size of the subset used to fit the trees at time $t$, and $\gamma=2$. A detailed explanation of its use with \texttt{R} code is provided in Supplementary material (see also the online notebook that allows for updates in the future). Stratification and smoothing over treatment history were allowed for static and dynamic rules, respectively. It did not find any positivity violation at $t=1$. We did not check positivity further for the first static rule of immediate initiation because the observed treatment is monotone (ART is not stopped once initiated, as illustrated in Figure \ref{alluviate}). No violation was identified for the never initiate ART static rule.

Positivity was challenged for initiating treatment under the two dynamic rules starting from six months (Tables \ref{tab:rule3} and \ref{tab:rule4}). The third rule of deferring ART with higher thresholds presented two violations at six months and thirteen violations at nine months. The fourth rule, deferring ART with lower thresholds, presented only one and seven violations at months 6 and 9, respectively. In contrast, no positivity violation was identified with smoothing over time, regardless of the rule. Many of the detected violations highlight the fact that it was unlikely for very healthy children to initiate ART immediately after becoming formally eligible according to older treatment rules. \textcolor{black}{The subgroups identified can be directly interpreted. Consider for instance the first violation provided in Table \ref{tab:rule3}. Its interpretation is that among individuals that did not initiate ART yet but should initiate at six months according to the third rule, and had \textit{both} a baseline CD4 count below 1600 and a baseline CD4 percentage greater or equal to 30\%, only 2.3\% actually initiated ART at six months. This constitutes a population with a rapid decline in their health status as treatment eligibility only occurs for CD4 counts <750 cells/mm$^2$ or CD\%$<25$\%. This subgroup represents 43 individuals or 5.4\% of those that should initiate ART at six months according to the rule. The prevalence of initiators in this subgroup is lower than the optimal positivity bound of 2.7\% computed from Gruber et al.'s formula \cite{Gruber_2022}. In contrast, the cumulative fitted propensity scores, as previously suggested,\cite{Cole_Hernan_2008, Petersen_2012} were less informative about the positivity violations. For instance, the nine-month cumulative estimates ranged from 0.0032 to 0.7152 for the third rule without stratification. Small values indicate positivity near-violations, but do not help the investigator to understand their underlying source.}    
 
These positivity violations may occur for three reasons, each with a different implication for the analysis. First, sparsity is more frequent in longitudinal settings due to censoring. Furthermore, the monotone treatment pattern reduces the number of untreated individuals over time. For instance, there were only 12 untreated individuals eligible for the fourth rule at nine months with baseline CD4\% lower than 15\% and a CD4 count between 100 and 225. In the case of sparsity, smoothing or using estimators able to extrapolate in the missing strata should be considered. Second, physicians may have deferred ART initiation for healthier children. Positivity violations will exist when the definition of “healthier” differs from those used in the intervention rule. In our data, ART was initiated according to the WHO 2010 guideline (i.e., the third rule),\cite{WHO_2010} which can explain some other rules' violations. Careful thinking about the definition of the intervention rules, and making sure that they fall within the variability of clinical practice at the time of data collection, will help to avoid them. Last, structural positivity violations may be present. For instance, HIV-positive pediatric ART coverage was estimated at around 33\% in Southern Africa in 2013\cite{Rosen_2023} and some unmeasured factors, such as socioeconomic status or nutrition, can lead to or be related to very low rates of ART initiation in some subgroups. Data quality and expert knowledge are crucial for this last case. \textcolor{black}{Since sPoRT is agnostic to the violation's nature, we must use our expert knowledge to judge whether the identified violation is likely structural or practical by discussing if there is a plausible underlying reason for this violation.\cite{Zhu_2021}}

\begin{table}[htb]
\begin{center}
\caption{Positivity violations identified with the stratified sPoRT for the third rule (dynamic): deferring ART with higher thresholds \label{tab:rule3}}%
\begin{tabular}{p{9cm}p{3cm}p{3cm}}
\hline
Subgroup & Prob. & \textit{n*} (\%) \\
\hline
\multicolumn{3}{l}{$P(A_t=1 | A_{t-1}=0, L_t, W, d_t(L_t)=1)$}\\
\multicolumn{3}{l}{\textbf{One month} ($n_1=2060$, $\beta=0.014$)}\\
 No violation was found & -  & -\\
 &&\\
\multicolumn{3}{l}{\textbf{Three months} ($n_2=1346$, $\beta=0.019$)}\\
No violation was found & - & - \\
&&\\
\multicolumn{3}{l}{\textbf{Six months} ($n_3=794$, $\beta=0.027$)}\\
CD4c$_0 < 1600$ \& CD4$\%_0 \geq 30$ & 0.023 & 43 (5.4) \\
CD4c$_3 < 2025$ \& CD4$\%_0 \geq 30$ & 0.025 & 40 (5.0) \\
&&\\
\multicolumn{3}{l}{\textbf{Nine months} ($n_4=565$, $\beta=0.033$)}\\
CD4$\%_4 \geq 25$ \& age $ = 4$ & 0.029 & 34 (6.0)\\
CD4$\%_4 \in \rinterval{25}{30}$ \& age $ < 3$ & 0.031 & 32 (5.7)\\
  CD4$\%_4 \geq 25$ \& CD4c$_0 \in \rinterval{625}{900}$ & 0.020 & 50 (8.8)\\
 CD4$\%_4 \in \rinterval{20}{25}$ \& CD4c$_0 \in \rinterval{900}{1225}$ & 0.000& 35 (6.2)\\
 HAZ$_0 \geq -1$ \& CD4$\%_0 \geq 25$ & 0.032 & 31 (5.5)\\
 CD4c$_4 \in \rinterval{625}{900}$ \& CD4$\%_0 \in \rinterval{25}{35}$ & 0.024 & 42 (7.4)\\
 CD4c$_4 \in \rinterval{1225}{1600}$ \& CD4$\%_0 \in \rinterval{10}{25}$ &  0.030& 33 (5.8)\\
 HAZ$_0 \in \rinterval{-4}{-1}$ \& WAZ$_0 \geq 0$ & 0.024 & 41 (7.3)\\
 CD4$\%_4 \in \rinterval{25}{30}$ \& CD4$\%_0 \geq 25$ & 0.024& 42 (7.4)\\
 CD4c$_4 \in \rinterval{1225}{1600}$ \& WAZ$_0 \geq -2$ & 0.030& 33 (5.8) \\
 CD4$\%_4 \geq 25$ \& WAZ$_0 \in \rinterval{-2}{0}$ & 0.032 & 63 (11.2)\\
 CD4$\%_4 \in \rinterval{20}{25}$ \& WAZ$_0 \geq 0$ & 0.000 & 29 (5.1)\\
 CD4$\%_4 \geq 25$ \& HAZ$_0 \in \rinterval{-2}{1}$ & 0.000 & 33 (5.8)\\
 & & \\
 \multicolumn{3}{l}{$P(A_t=0 | A_{t-1}=0, L_t, W, d_t(L_t)=0)$}\\
 \multicolumn{3}{l}{\textbf{One month} ($n_1=292$, $\beta=0.052$)}\\
 No violation was found & -  & -\\
 &&\\
 \multicolumn{3}{l}{\textbf{Three months} ($n_2=206$, $\beta=0.065$)}\\
 No violation was found & - & -\\
& & \\
\multicolumn{3}{l}{\textbf{Six months} ($n_3=137$, $\beta=0.087$)}\\
No violation was found & - & -\\
& & \\
\multicolumn{3}{l}{\textbf{Nine months} ($n_4=112$, $\beta=0.100$)}\\
No violation was found & - & -\\
\hline
\multicolumn{3}{p{15cm}}{Abbreviations: $\beta$, Gruber bound; CD4c, CD4 count; HAZ, height for age z-score; $n_t$, sample size at time $t$; \textit{n*}, subgroup size; Prob., probability of being treated; WAZ, weight for age z-score. Subscript indicates time-point. Smoothing over treatment history was considered.}
\end{tabular}
\end{center}
\end{table}

\begin{table}[htb]
\begin{center}
\caption{Positivity violations identified with the stratified sPoRT for the fourth rule (dynamic): deferring ART with lower thresholds \label{tab:rule4}}%
\begin{tabular}{p{9cm}p{3cm}p{3cm}}
\hline
Subgroup & Prob. & \textit{n*} (\%) \\
\hline
\multicolumn{3}{l}{$P(A_t=1 | A_{t-1}=0, L_t, W, d_t(L_t)=1)$}\\
\multicolumn{3}{l}{\textbf{One month} ($n_1=1164$, $\beta=0.021$)}\\
 No violation was found & -  & -\\
 &&\\
\multicolumn{3}{l}{\textbf{Three months} ($n_2=720$, $\beta=0.028$)}\\
 No violation was found & -  & -\\
&&\\
\multicolumn{3}{l}{\textbf{Six months} ($n_3=354$, $\beta=0.045$)}\\
HAZ$_0 \geq -1$ \& Male & 0.038  & 26 (7.3) \\
&&\\
\multicolumn{3}{l}{\textbf{Nine months} ($n_4=226$, $\beta=0.061$)}\\
WAZ$_4 \in \rinterval{-2}{-1}$ \& age $ \geq 3$ & 0.048  & 42 (18.6) \\
 CD4c$_0 \in \rinterval{100}{225}$ \& CD4$\%_0 < 15$ & 0.000 & 12 (5.3) \\
  CD4c$_4 < 400$ \& CD4c$_0 \in \rinterval{100}{225}$ & 0.000 & 12 (5.3) \\
 CD4c$_0 < 400$ \& WAZ$_4 \in \rinterval{-2}{-1}$ & 0.034 & 26 (12.8)\\
 CD4c$_4 \in \rinterval{100}{225}$ \& CD4$\%_0 < 15$ & 0.000  & 15 (6.6)\\
 WAZ$_4 \in \rinterval{-2}{-1}$ \& CD4$\%_0 \in \rinterval{10}{15}$ &  0.043 & 23 (10.2) \\
 CD4c$_4 \in \rinterval{100}{400}$ \& HAZ$_0 \in \rinterval{-3}{-2}$ & 0.056 & 18 (8.0) \\
 & & \\
 \multicolumn{3}{l}{$P(A_t=0 | A_{t-1}=0, L_t, W, d_t(L_t)=0)$}\\
 \multicolumn{3}{l}{\textbf{One month} ($n_1=1188$, $\beta=0.020$)}\\
 No violation was found & -  & -\\
 &&\\
 \multicolumn{3}{l}{\textbf{Three months} ($n_2=832$, $\beta=0.026$)}\\
  No violation was found & - & - \\
&&\\\multicolumn{3}{l}{\textbf{Six months} ($n_3=577$, $\beta=0.033$)}\\
   No violation was found & - & - \\
&&\\
  \multicolumn{3}{l}{\textbf{Nine months} ($n_4=451$, $\beta=0.039$)}\\
  No violation was found & - & - \\
\hline
\multicolumn{3}{p{15cm}}{Abbreviations: $\beta$, Gruber bound; CD4c, CD4 count; HAZ, height for age z-score; $n_t$, sample size at time $t$; \textit{n*}, subgroup size; Prob., probability of being treated; WAZ, weight for age z-score. Subscript indicates time-point.  Smoothing over treatment history was considered.} 
\end{tabular}
\end{center}
\end{table}

\textcolor{black}{Finally, sPoRT can also be used to check positivity violations involved in the censoring process.\cite{Hernan_Robins_2020} In such a case, the dependent variable is the censoring status, and the treatment was added to the adjustment set. Note that the treatment rule plays no role here. No violation was identified in our application, regardless of the smoothing level.}

\section{Discussion}

In this paper, we described sPoRT, a new algorithm for checking sequential positivity by identifying possibly overlapping subgroups of individuals without support for the defined intervention strategies. The algorithm can match the desired level of smoothing in the analysis. We implemented the algorithm in the \texttt{R} package \texttt{PoRT} to facilitate its use and provide an \texttt{R} notebook demonstrating its application. \textcolor{black}{While we only presented a case study, not directly generalizable to all situations, the following recommendations apply more broadly.}

The standard way to check sequential positivity is to compute several propensity scores and use the cumulative product of their resulting weights as a measure of data support, a tiny cumulative weight being a red flag for non-positivity.\cite{Schomaker_2019, Petersen_Schwab_2014} However, this approach  (i) is based on the assumption that all the propensity scores usually fitted using parametric regressions are correctly specified and (ii) does not directly identify problematic subgroups of individuals. Indeed, although individuals with near-zero predicted probabilities of following strategies are easily identifiable, their shared characteristics differing from the remaining individuals may not be obvious. In contrast, sPoRT provides subgroups defined by the values of one or more covariates, possibly at each time point when stratification on time is desired. The usage of regression trees avoids the parametric assumptions of modeling the propensity scores, adding robustness to the sequential positivity checking. We recommend the routine usage of both approaches. \textcolor{black}{Note that the estimated cumulative propensity scores presented in our application involved both treatment and censoring models to deal with artificial selection bias.\cite{Hernan_2016} Additionally, a bootstrap diagnostic tool estimating the bias related to a lack of positivity was developed for cross-sectional settings,\cite{Petersen_2012} and adapted to not rely on treatment simulation.\cite{Bahamyirou_2019} Its extension to longitudinal settings would be a useful addition to the epidemiologists' toolbox.}

However, we stress that both the fitting of MSMs and the usage of sPoRT require the correct a priori identification of the confounders and time-ordering of all variables. Only domain knowledge can inform this. Furthermore, sPoRT depends on its tuning parameters $\alpha$, $\beta$ and $\gamma$ (and those specific to the regression tree to a lesser extent). A grid of pragmatically informed values (as suggested in Table \ref{tab:port}) can be considered for these parameters.\cite{Danelian_2023} Unlike Danelian \textit{et al.},\cite{Danelian_2023} we adapted $\beta$ automatically to the sample size at a given time using Gruber's bound.\cite{Gruber_2022} Since the bound is a function of the sample size, it naturally accounts for the decreasing sample size over time in longitudinal contexts. However, this \textcolor{black}{optimal (in terms of mean square error)} bound was initially defined for cross-sectional studies; thereby, the optimal bound function could differ in longitudinal settings. \textcolor{black}{Note that when the adjustment set is high-dimensional, a trade-off should be found between non-positivity and residual confounding by excluding weak confounders (i.e., those having an expected strong association with exposure and a weak one with the outcome).\cite{Cole_Hernan_2008}}

\textcolor{black}{In our application, we also used sPoRT to check positivity with respect to censoring in order to identify subgroups of individuals that can never be censored. While we did not identify such subgroups in this particular analysis, some subtle differences compared to exposure positivity merit consideration. First, censoring positivity must be checked only for the never-censored, since we are interested in the counterfactual outcomes under no censoring. Therefore, one falls into a static rule with a monotone pattern of censoring, and we should check whether $\mathrm{P}(C_t=1|C_{t-1}=0, X_t=x_t)>\beta\geq 0$, where $C$ is the censoring indicator (1 if censored and 0 otherwise) and $X_t=x_t$ a particular subgroup defined by the adjustement set. The adjustment set may also not overlap entirely between the exposure and censoring sequential conditional exchangeability, and needlessly combining these respective covariates sets should be avoided \cite{Zivich_2022}. Similarly to confounding, directed acyclic graphs can be used for selecting covariates used to control such selection bias \cite{Hernan_Hernandez_2004, Howe_2016}. The values of the hyperparameters $\alpha$ and $\beta$ might also be adapted according to the clinical question at hand and the specific settings. For instance, $\beta$'s value could be increased with the censoring rate, since the prevalence of the sPoRT's dependent variable becomes more balanced. The use of Gruber's bound could also be questioned, and its optimality in this setting should be investigated in the future. While structural violations yield a redefinition of the rule, the amount of practical violations is also of interest when inverse probability of censoring weighting is considered in the analysis to control for selection bias.}

Finally, what should be done with the sPoRT output? We recommend checking first whether some patterns regularly appear over time and the subgroups from the first time point to consider potential structural positivity violations \textcolor{black}{that make the initial question unanswerable}. If some subgroups are likely due to structural violations at baseline, they must be excluded by adding exclusion criteria to the study, and then positivity should be rechecked to see if other violations on the variables involved in these subgroups were hidden. \textcolor{black}{Post-baseline s}tructural positivity violations \textcolor{black}{must be handled} by adapting the intervention strategy to account for these violations (through the addition of an ``inability to treat'' criterion).\cite{Schnitzer_Platt_Durand_2020, Jensen_2024} Once the structural violations have been ruled out, sparsity can be investigated through the remaining subgroups. The violations and subgroup sizes may influence the choice of the estimator. For instance, inverse probability weighting is highly sensitive to sparsity.\cite{Spreafico_2024} In contrast, estimators able to extrapolate over missing strata could be considered, such as the g-formula.\cite{Robins_2007} When the sparsity is high, alternative estimands could also be considered; see Hoffman \textit{et al.} for an overview.\cite{Hoffman_2024} \textcolor{black}{In our motivating example, we would ideally recommend measuring a proxy of health-care accessibility before conducting the analysis to investigate potential structural violations. More pragmatically, propensity scores smoothing over time yielded no practical positivity violations, regardless of the rule, and could be preferred over time-stratified propensity scores in this application, as we relied on short-term treatment and confounder histories.}

\newpage
    
\textbf{Conflict of interest statement:} None declared.

\textbf{Data statement:}
Data ownership remains at the clinic level, and the IeDEA regional steering groups must approve any analyses. Data access requests can be sent through https://www.iedea-sa.org/collaborate-with-us/.

\textbf{Author contributions statement:}
AC: Conceptualization, Methodology, Software, Writing - Original Draft, and Writing - Review \& Editing. MS: Data curation, Methodology, and Writing - Review \& Editing. MALF: Writing - Review \& Editing. RWP: Supervision and Writing - Review \& Editing. MES: Methodology, Supervision, and Writing - Review \& Editing.

\textbf{Funding:}
A.C. was supported by an IVADO postdoctoral fellowship (\#2022-7820036733). MES holds a tier 2 Canada Research Chair in Causal Inference and Machine Learning in Health Science and a Discovery Grant (\#RGPIN-2021-03019) from the Natural Sciences and Engineering Research Council of Canada. 

\textbf{Acknowledgement:} This project was approved by the Universit\'e de Montr\'eal clinical ethics committee (\#2021-359). We also thank Shobna Sawry, Karl-G{\"u}nter Technau, Frank Tanser, Haby Sam Phiri, Vivian Cox, Cleophas Chimbete, Janet Giddy, and Robin Wood for sharing their data with us. We would also like to highlight the support of the International epidemiologic Databases to Evaluate AIDS Southern Africa (IeDEA-SA), led by Matthias Egger and Mary-Ann Davies; the NIH has supported the collaboration, grant numbers \#U01AI069924. 

\textbf{Publication history:} A preprint was published on arXiv (doi: 10.48550/arXiv.2412.10245).

\newpage


\bibliographystyle{vancouver}
\bibliography{reference}


\end{document}